\documentclass[aps,pre,twocolumn,superscriptaddress]{revtex4-1}  

\usepackage{color}
\usepackage{amsmath}      	
\usepackage{amssymb} 
\usepackage{amsfonts}   
\usepackage{graphicx}
\usepackage{bm}
\usepackage{ifthen}
\newboolean{pnas}
\setboolean{pnas}{false}

\usepackage{bbm}
\usepackage{textcomp}

\newcommand{\eq}[1]{\begin{align} #1 \end{align}}
		
\newcommand{\<}{\langle}
\newcommand{\beq}{\begin{equation}}
\newcommand{\eeq}{\end{equation}}
\renewcommand{\>}{\rangle} 

\newcommand{\s}{\sigma}

\newcommand{\cffig}{Fig.~}
\newcommand{\cfeq}{eq.~}
\newcommand{\cfMM}{{Materials and Methods}}
\newcommand{\cfSI}{data not shown}

\makeatletter
\def\@seccntformat#1{%
  \expandafter\ifx\csname c@#1\endcsname\c@section\else
  \csname the#1\endcsname\quad
  \fi}
\makeatother

\begin{document}

\title{Blindfold learning of an accurate neural metric}

\author{Christophe Gardella}

\affiliation{Laboratoire de physique statistique, CNRS, UPMC, Universit\'e Paris Diderot, and \'Ecole normale sup\'erieure (PSL Research University), 24 rue Lhomond, 75005 Paris, France}
\affiliation{Institut de la Vision, INSERM and UMPC, 17 rue Moreau, 75012 Paris, France}
\author{Olivier Marre}
\thanks{These authors contributed equally. Correspondence should be sent to \url{olivier.marre@inserm.fr} and \url{tmora@lps.ens.fr}.}
\affiliation{Institut de la Vision, INSERM and UMPC, 17 rue Moreau, 75012 Paris, France}
\author{Thierry Mora}
\thanks{These authors contributed equally. Correspondence should be sent to \url{olivier.marre@inserm.fr} and \url{tmora@lps.ens.fr}.}
\affiliation{Laboratoire de physique statistique, CNRS, UPMC, Universit\'e Paris Diderot, and \'Ecole normale sup\'erieure (PSL Research University), 24 rue Lhomond, 75005 Paris, France}

\begin{abstract}
The brain has no direct access to physical stimuli, but only to the spiking activity evoked in sensory organs. It is unclear how the brain can structure its representation of the world based on differences between those noisy, correlated responses alone. Here we show how to build a distance map of responses from the structure of the population activity of retinal ganglion cells, allowing for the accurate discrimination of distinct visual stimuli from the retinal response. We introduce the Temporal Restricted Boltzmann Machine to learn the spatiotemporal structure of the population activity, and use this model to define a distance between spike trains. We show that this metric outperforms existing neural distances at discriminating pairs of stimuli that are barely distinguishable. The proposed method provides a generic and biologically plausible way to learn to associate similar stimuli based on their spiking responses, without any other knowledge of these stimuli.

\end{abstract}

\maketitle


A major challenge in neuroscience is to understand how the brain processes sensory stimuli. 
In particular, the brain must learn to group some stimuli in the same category, and to discriminate others.
Strikingly, this feat is achieved while the brain has only access to the noisy responses evoked in sensory organs, but never to the stimulus itself. For example, the brain only receives the retinal responses to visual stimuli, and is able to associate together responses corresponding to the same stimulus, while teasing apart the ones coming from distinguishable stimuli. How nervous systems can achieve such discrimination is still unclear. 
One strategy to solve this problem could be to learn either a decoding model to reconstruct the stimulus from the neural responses \citep{Warland1997,Marre2015}, or an encoding model and invert it to find stimuli that can be distinguished \cite{Ganmor2015}. However, in both cases, this requires to have access to a lot of pairs of stimuli and evoked responses. Clearly, the brain is not guaranteed to have access to such data, and may only access the neural response without knowing the corresponding stimulus. 

Neural metrics, which define a distance between pairs of spike trains, have been proposed to solve this issue. In general, spike trains evoked by the same stimulus should be close by, while spike trains corresponding to very different stimuli should be far away. Using a given metric, one can associate together responses evoked by similar stimuli, without any information about the stimuli themselves \citep{Machens2003,Narayan2006}. The quality of this classification relies on the metric being well adapted to the task at hand, and different metrics are not expected to perform equally well.

Multiple metrics based on different features of the neural response have been proposed, mostly for single cells \citep{Victor1996,Victor1996,Berry1997,VanRossum2001,Quiroga2002,Schreiber2003,Hunter2003}, and exceptionally for populations \cite{Houghton2008}. These metrics do not use information about the correlative structure of the population response, and often require to tune parameters to optimize performance, which requires external knowledge of the stimulus. 
In addition, a precise quantification of the performance of these different metrics at discriminating barely distinguishable stimuli is lacking.

Here we present an approach to learn a spike train metric with high discrimination capacity from the statistical structure of the population activity itself. We applied the method to the retina, a sensory system characterized by noisy, non-linear \citep{Gollisch2010}, and correlated \citep{Arnett1978,Schneidman2006} responses.
We first introduce a statistical model of retinal responses, the Temporal Restricted Boltzmann Machine, which allows us to learn an accurate description of spatio-temporal correlations in a population of 60 ganglion cells of the rat retina, stimulated by a randomly moving bar.
We then use this model to derive a metric on neural responses. Using closed-loop experiments, where stimuli are tuned to be hardly distinguishable from each other, we show that this neural metric outperforms classical metrics at stimulus discrimination tasks. This high discrimination capacity is achieved despite the neural metric being trained with no information about the stimulus.
We therefore suggest a general and biologically realistic method for the brain to learn to efficiently discriminate stimuli solely based on the output of sensory organs. 

\section*{Results}

\begin{figure}
\begin{center}
\includegraphics[width=\linewidth]{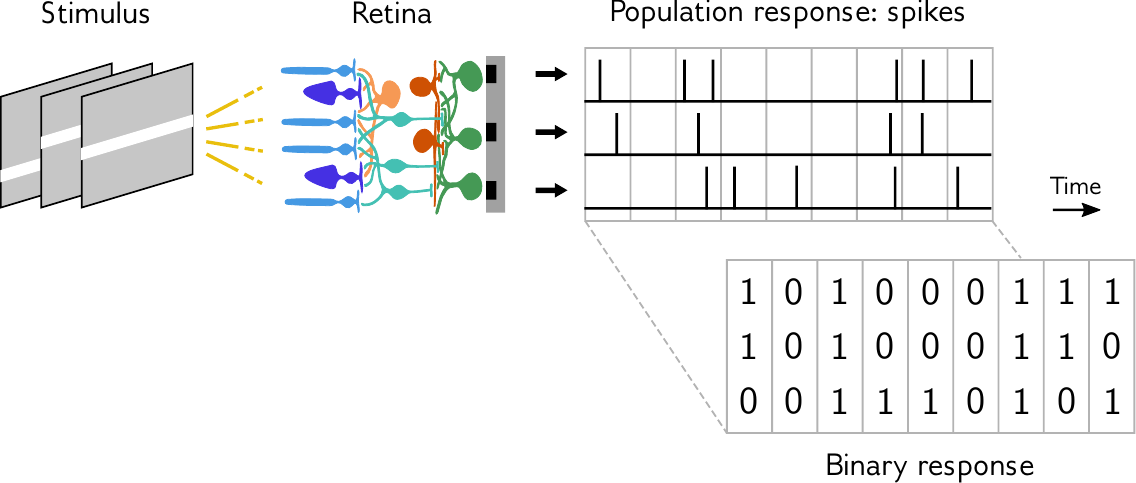}
\end{center}
\caption{
Experimental setup.  
A rat retina is stimulated with a moving bar. Retinal ganglion cells
(in green) are recorded with a multielectrode array. To model
the response, spike trains are binarized in 20 ms time bins. 
}
\label{f:setup}
\end{figure}

\subsection*{Modeling synchronous population activity with Restricted Boltzmann Machines}
We analyzed previously published \textit{ex vivo} recordings from rat retinal ganglion cells \citep{Ferrari2016closed}. 
A population of 60 cells was stimulated with a moving bar and recorded with a multielectrode array (\cffig\ref{f:setup}). Responses were binarized in 20 ms time bins, with value $\sigma_{it}=1$ if neuron $i$ spiked during in given time bin $t$, and 0 otherwise (\cffig\ref{f:setup}). We first aimed to describe the collective statistics of spikes and silences in the retinal population, with no regard for the sequence of stimuli that evoked them.

\begin{figure} 
\begin{center}
\includegraphics[width=\linewidth]{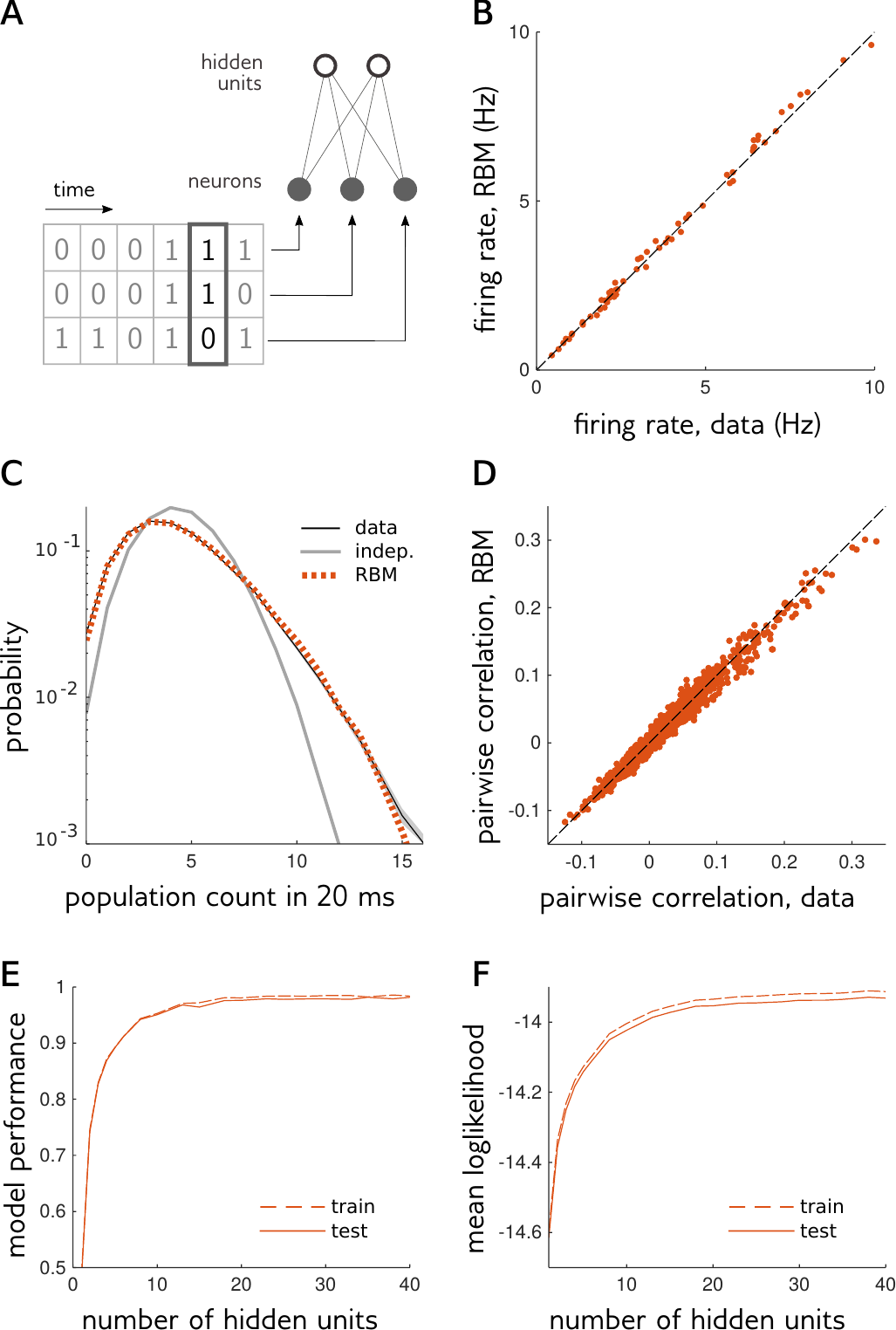}
\end{center}
\caption{
The Restricted Boltzmann Machine (RBM) model predicts accurately response statistics within single 20 ms time bins. 
{\textbf A}, The RBM models the probability of binarized responses in single time bins. There is no direct interactions between neurons (grey circles). Instead, neurons interact with hidden units (white circles).
{\textbf B}, Single cell firing rate. Each dot represents the spiking frequency of a neuron in the testing set (not used for learning), versus RBM model prediction. 
{\textbf C}, Distribution of the total number of spikes in the population during a time bin in the testing set (black) versus the prediction of a model of independent neurons (gray), or by the RBM (dotted red). Shaded area shows standard error in data.
{\textbf D}, Pairwise correlations. Each dot represents the Pearson correlation for a pair of neurons, in the testing set versus RBM prediction.
{\textbf E}, Fraction of the variance of correlations explained by RBM models, for different numbers of hidden units, in the training and testing sets.
{\bf F},  Mean model log-likelihood in-sample (dashed line) and out-of-sample (full line) as a function of the number of hidden units. The small difference between training and testing sets suggests that there is no over-fitting. 
}
\label{f:stats_RBM}
\end{figure}

We modeled synchronous correlations between neurons using Restricted Boltzmann Machines (RBMs) \cite{Smolensky1986,Hinton2002}, which have previously been applied to retinal \cite{Schwab2013,Humplik2016} and cortical \cite{Koster2014} populations. They give the probability of same-time spikewords $(\sigma_{i})=(\sigma_{it})_i$ at any $t$ as:
\beq
P[(\sigma_{i})]=\frac{1}{Z}\sum_{(h_j)}\exp\left(\sum_{i} a_i \s_{i} + \sum_{j} b_j h_{j} + \sum_{i,j} W_{ji} \s_{i} h_{j}\right)
\eeq
RBMs do not have direct interactions between neurons. Rather, their correlations are explained by interactions with {binary} latent variables, $h_j$, called hidden units (\cffig\ref{f:stats_RBM}A). 
When a hidden unit takes value 1, it induces collective changes in the excitability of sub-populations of cells.
Although it is tempting to think of hidden units as non-visible neurons, they are only effective variables and usually do not correspond to actual neurons;
hidden units can in fact reflect multiple causes of correlations, such as direct input from neighboring cells, or common input from intermediate layers and the stimulus. Their number can be varied: the more hidden units, the more complex structures can be reproduced by the model, but the more parameters need to be estimated.

We learned an RBM with 20 hidden units to model the responses of the retinal population responding to a randomly moving bar.
The model was inferred on a training set (80\% of responses) using persistent contrastive divergence (\cfMM), and its
predictions compared to a testing set (20\% remaining responses).
The RBM predicted well each neuron's firing rate (\cffig\ref{f:stats_RBM}B) as well as correlations between pairs of neurons (\cffig\ref{f:stats_RBM}D).
In addition, the RBM predicted higher-order correlations accurately, such as the distribution of the total number of spikes in the population (\cffig\ref{f:stats_RBM}C).
By contrast, a model of independent neurons (zero hidden units) under-estimated the probability of events with few or many spikes by an order of magnitude.
The model performance, measured by either the fraction of variance explained of pairwise correlations (\cffig\ref{f:stats_RBM}E), or by the model log-likelihood (\cffig\ref{f:stats_RBM}F), quickly saturated with the number of hidden units, with 15 units already providing near optimal performance.

\begin{figure}[!htbp]
\centering
\includegraphics[width=0.9\linewidth]{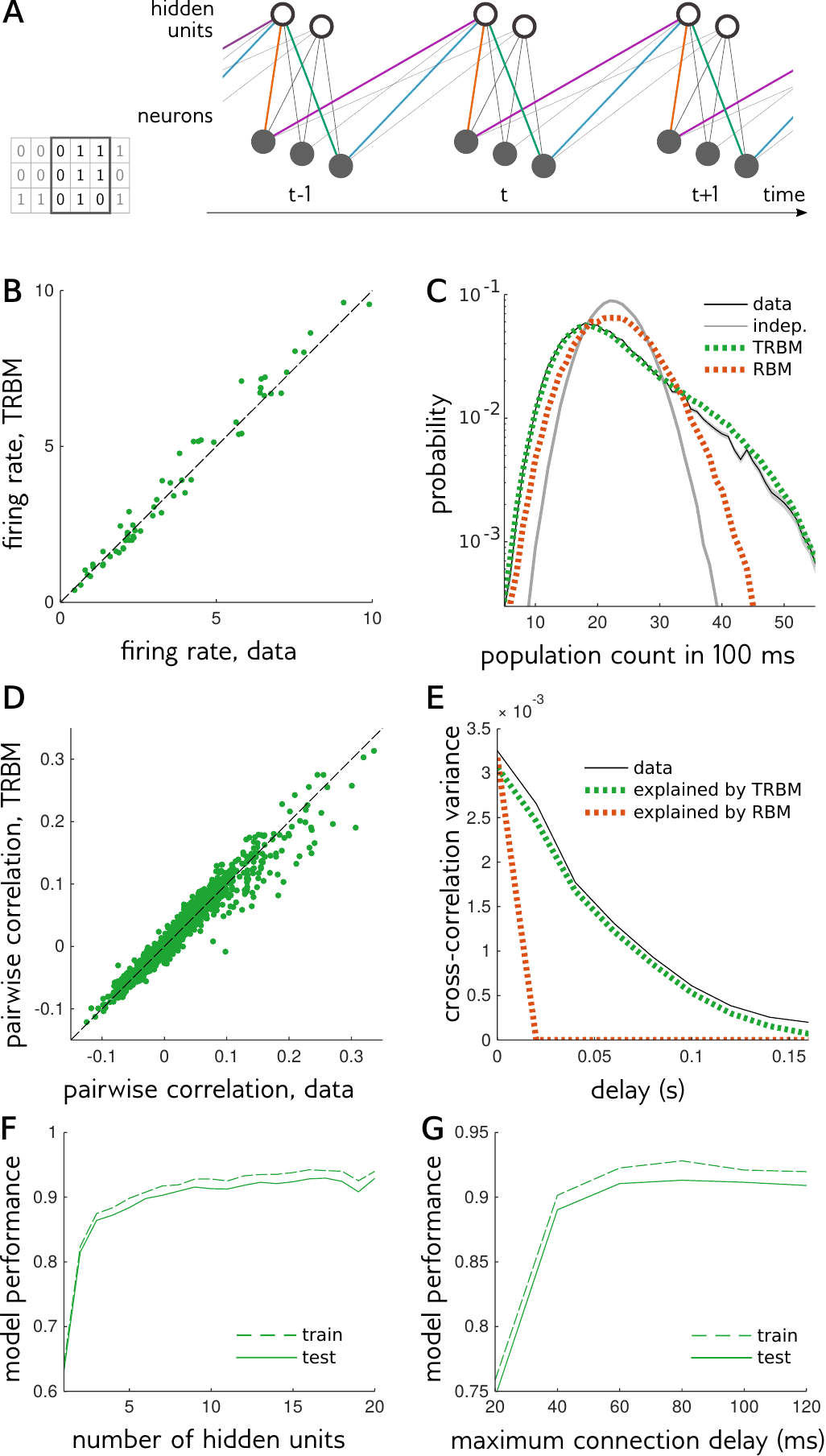}
\caption{
The Temporal Restricted Boltzmann Machine (TRBM) model predicts accurate response statistics across multiple time bins. 
{\textbf A}, The TRBM's structure is similar to the RBM's, but neurons and hidden units are connected across multiple time bins. The interaction between neurons and hidden units only depends on the delay between them: in this schematic interactions with the same color are equal. For simplicity, the model represented here only has interactions for delay 0 and 1 time bins. In general there can be interactions with larger time delays. 
{\textbf B}, Single cell firing rates. Same as \cffig\ref{f:stats_RBM}B but for TRBM model.
{\textbf C}, Distribution of the number of spikes in the population during a 100~ms time window (5 consecutive time bins), in the testing set (black), predicted by a model of independent time bins and independent neurons (grey), a model with independent RBMs in each time bin (dotted red), or a TRBM (dotted green).  Shaded area shows standard error in data.
{\textbf D}, Pairwise correlation. Same as \cffig\ref{f:stats_RBM}D but for TRBM model. 
{\textbf E}, Cross-correlation. Black line show the variance in cross-correlations between neurons with different time delays. Red and green lines show variance explained by RBM and TRBM respectively.
{\textbf F}, Fraction of the variance of cross-correlations between neurons with delays up to 140 ms explained by TRBM models, as a function of the number of hidden units, in the training and the testing sets.
{\textbf G}, Same as F, but varying the maximum connection delay between hidden and visible units.
}
\label{f:stats_TRBM}
\end{figure}

\subsection*{Temporal Restricted Boltzmann Machines for population spike trains}	
The RBM performs well at modeling neural responses within 20~ms time bins, but correlations between neurons often span longer time scales. 
To evaluate the importance of these longer term correlations, we plotted the distribution of the number of spikes in the population in a 100~ms time window (using the testing set), and compared it to the prediction from the RBM, where the response of the population was generated in each of the five 20-ms bin independently
(\cffig\ref{f:stats_TRBM}C).
Although the RBM performed better than a model of independent neurons, it still underestimated the probability of large numbers of spikes by an order of magnitude, indicating that correlations over longer scales than 20 ms play an important role in shaping the collective response statistics.

To account for these temporal correlations, we introduced the Temporal Restricted Boltzmann Machine (TRBM). This model generalizes the RBM by allowing for interactions between neurons and hidden units across different time bins (\cffig\ref{f:stats_TRBM}A, \cfMM):
\beq
\begin{split}
P[(\sigma_{it})]=&\frac{1}{Z}\sum_{(h_{jt'})}\exp\left(\sum_{it} a_i \s_{it} + \sum_{jt'} b_j h_{jt'} \right.\\&\left.+ \sum_{i,j,t,t'} W_{ji,{t'-t}} \s_{it} h_{jt'}\right).
\end{split}
\eeq
Because we want to describe the stationary distribution of spike trains regardless of the stimulus, absolute time is irrelevant, and
the model is invariant to time translations: {connections between a hidden unit and a neuron only depend on the relative delay $t'-t$ between them}. This property is similar to convolutional networks used in image processing, but here in time instead of space.

We trained a TRBM with 10 hidden units per time bin, each connected to neurons across 5 consecutive time bins, on the same training set as before using persistent contrastive divergence (\cfMM), and compared predictions to the testing set.
Like the RBM, the TRBM could predict individual neuron firing rates (\cffig\ref{f:stats_TRBM}B) and synchronous pairwise correlations (\cffig\ref{f:stats_TRBM}D).
In addition, the TRBM could also predict temporal correlations ignored by the RBM.
In particular it reproduced accurately the distribution of the total number of spikes in a 100~ms time window, that the RBM did not (\cffig\ref{f:stats_TRBM}C).
We also tested if the TRBM could predict correlations between the spiking activity of pairs of neurons in two time bins separated by a given delay.
{To do so, we computed the total variance of pairwise correlations for a each delay, and estimated the fraction of it that could be explained by the TRBM (\cffig\ref{f:stats_TRBM}E, \cfMM).}
Even though direct connections between neurons and hidden units were limited to 80 ms, the TRBM could explain a substantial amount of correlations even for large delays, up to 150~ms where correlations all but vanish. 

Similarly to the RBM, we found that increasing the number of hidden units only marginally improved performance (as measured by the fraction of explained variance of pairwise correlations) beyond 10 units per time bin (\cffig\ref{f:stats_TRBM}F).
We also varied the maximum connection delay between neurons and hidden units from 20~ms to 120~ms. Performance quickly saturated at a connection delay of around 60~ms (\cffig\ref{f:stats_TRBM}G).
In the following we will consider a TRBM with 10 hidden units and connection delay of 80~ms, unless mentioned otherwise.

\subsection*{A neural metric based on response statistics}

{The hidden units of the TRBM can be considered as a way to compress the variability present in the neural activity, and extract its most relevant dimensions. We asked whether these hidden units could be used to define a neural metric that would follow the structure of the population code,  allowing for efficient discrimination and classification properties.}

To this end, we designed neural metrics derived from the RBM and TRBM based on {the difference between the hidden unit states. 
Take two responses ${\boldsymbol \sigma}=(\sigma_{i})$ and ${\boldsymbol \sigma}'=(\sigma'_{i})$ of the retina, and define ${\boldsymbol \Delta\boldsymbol h}=(\Delta h_{j})$ as the difference of mean value of the hidden units conditioned on the two responses, $\Delta h_{j}=\<h_{j}\>_{\boldsymbol \sigma}-\<h_{j}\>_{\boldsymbol \sigma'}$ (\cfMM). Then the RBM metric is defined as:
\beq\label{eq:metric}
d_{\rm RBM}=\bm{\Delta h}^\intercal \bm{W C W}^\intercal \bm{\Delta h},
\eeq
where $\bm{C}=\<\bm{\sigma}\bm{\sigma}^\intercal\>-\<\bm{\sigma}\>\<\bm{\sigma}^\intercal\>$ is the covariance matrix of the response, and $\bm{W}=(W_{ji})$ is the matrix of couplings between neurons and hidden units. This definition can readily be generalized to the TRBM by adding time indices ({\cfMM}).
Note that this metric differs from the Euclidian distance in the space of hidden units, $\bm{\Delta h}^\intercal\bm{\Delta h}$: it has a nontrivial kernel $\bm{W C W}^\intercal$, which modulates the contribution of each hidden unit by its impact on neural activity. We will see later that this kernel improves discrimination capabilities.
{Note that this metric was defined without any information about the stimulus, and solely from the knowledge of the activity. We next aimed to test how well this metric can discriminate pairs of stimuli.}

\begin{figure}[!b]
\centering
\includegraphics[width=1\linewidth]{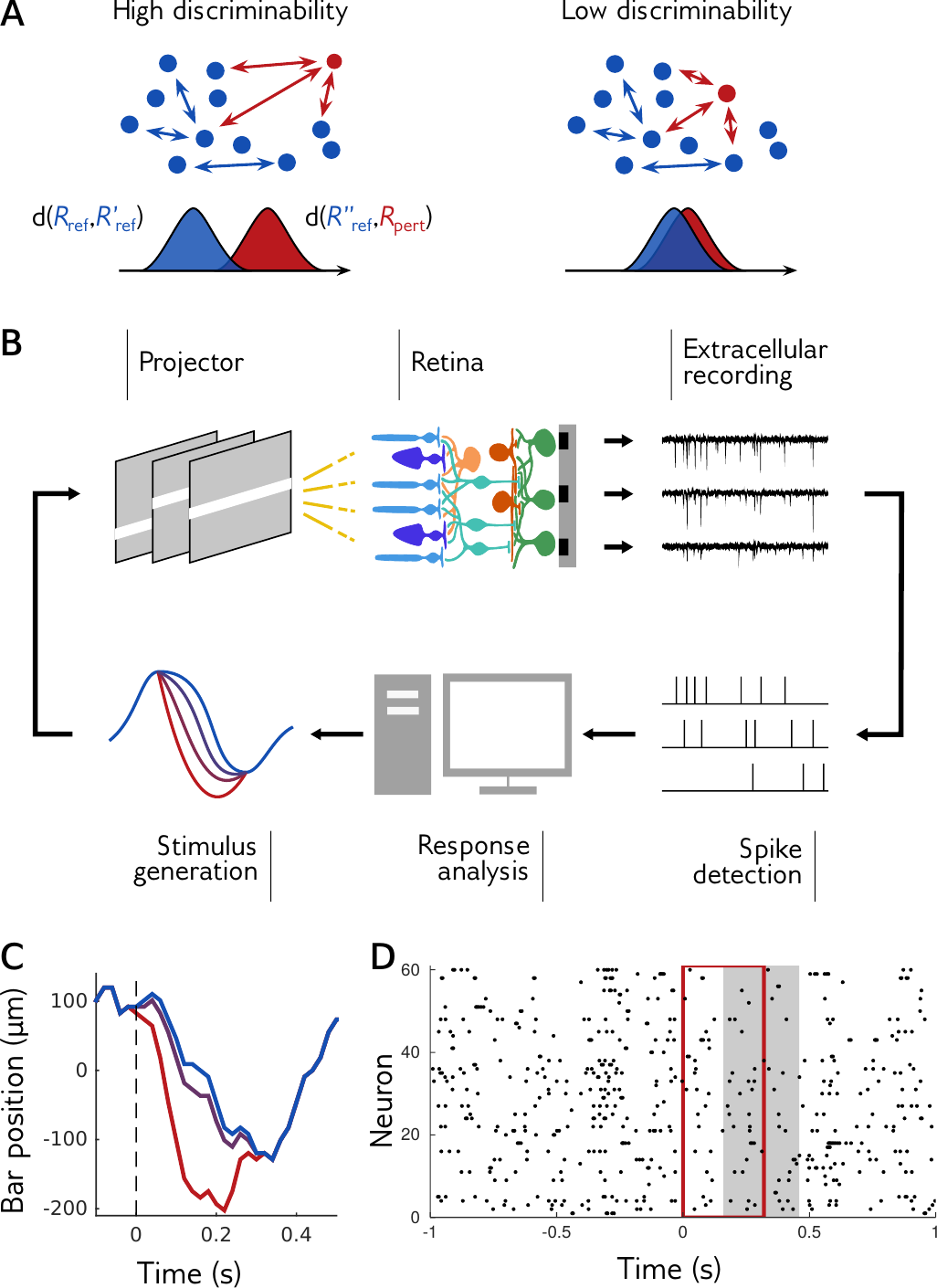}
\caption{
Online adaptation of perturbations.
{\bf A}, Discriminating with metrics. Stimulus discrimination is
evaluted by comparing the distance of responses within the same
reference stimulus (blue dots), and between the reference and a
perturbation (red dots). Discrimability is defined as the probability
that a within-stimulus distance (blue distribution) is lower than an
across-stimuli distance (red distribution).
{\textbf B}, Closed-loop experiment. 
At each step, the rat retina was stimulated with a perturbation of a reference stimulus. 
Retinal ganglion cell responses were recorded extracellularly with a multi-electrode array.
Electrode signals were high-pass filtered and spikes were detected by threshold crossing. 
We computed the discriminability of the population response, and adapted the amplitude of the next perturbation. 
{\textbf C}, 
The stimulus consisted in repetitions of a reference stimulus (here
the trajectory of a bar, in blue), and in perturbations of this
reference stimulus of different shapes and amplitudes (see Fig.~S1). Purple and red trajectories are perturbations with the same shape, at small and large amplitude.
{\textbf D}, Example population response. Each spike is represented by a dot. Red rectangle: duration of the perturbation. Shaded rectangle: duration of responses for which the discriminability was measured.
}
\label{f:closed_loop}
\end{figure}

\subsection*{Distinguishing close stimuli}
To evaluate the capacity of a neural metric {to} finely resolve stimuli based on the sensory response, we introduce a measure a discriminability between the responses to two distinct stimuli based on neural {metrics}. 

The response to a given stimulus is intrinsically noisy. Two repetitions of the same stimulus (let us call it reference stimulus) will give rise to two distinct responses $R_\text{ref}$ and $R'_\text{ref}$. 
The response $R_{\text{pert}}$ to a perturbation of the reference stimulus 
may thus be hard to tease apart from another response to the reference stimulus, because of this noise (\cffig\ref{f:discrim__Bar}A). 
Given a neural metric $d(R,R')$, it is natural to define the discriminability of a perturbation as the probability for the response $R_\text{pert}$ to be further apart from a response to the reference, $R''_\text{ref}$, than would two responses to the reference, $R_\text{ref}$ and $R'_\text{ref}$, from each other:
\beq\label{eq:discr}
{\rm Discr}=P(d(R''_{\text{ref}},R_{\text{pert}})>d(R_{\text{ref}},R'_{\text{ref}})).
\eeq
If a perturbation is perfectly discriminable (\cffig\ref{f:discrim__Bar}A, left), distances between reference and perturbation are well separated from distances within responses to the reference, and 
the discriminability will approach 1.
Conversely, for perturbations too small to be discriminated, the two distributions greatly overlap (\cffig\ref{f:discrim__Bar}A, right), and the discriminability is close to $0.5$ corresponding to chance.

To finely assess the capacity of neural metrics to perform discrimination tasks, we need to study perturbations that lie between these two extremes, where discrimination is neither easy nor impossible. To find this soft spot, we performed closed-loop experiments where at each step the discriminability of a perturbation was analyzed in order to generate the perturbation at the next step (\cffig\ref{f:closed_loop}B, see \cite{Ferrari2016closed} for more details). We first recorded multiple responses to a reference stimulus, a $0.9$\,s snippet of bar trajectory described earlier (Fig.~S1\, A-B). We then recorded responses to many perturbations of this stimulus (\cffig\ref{f:closed_loop}C). For a given ``shape'' of the perturbation (i.e. normalized difference of bar position between reference and perturbation as a function of time, Fig.~S1\, C), we adapted the perturbation size online, and searched for the smallest perturbations that were still discriminable (\cfMM). If a perturbation had high discriminability (as defined by a linear discrimination task on the thresholded values of the raw multi-electrode array output, independently of any metric, see {\cfMM}), at the next step we tested a perturbation with smaller amplitude.
Conversely, if a perturbation had low discriminability, we then tested a larger perturbation. Perturbations lasted 320~ms, and responses were analyzed over 300~ms with a delay (\cffig\ref{f:closed_loop}D).

Thanks to this method, we could explore the space of possible perturbations efficiently, exploring multiple directions (shapes) of the perturbation space simultaneously, and obtained a range of responses to pairs of stimuli that are challenging but not impossible to discriminate. This method allowed us to benchmark different metrics.

\begin{figure*}[!htbp]
\centering
\includegraphics[width=0.85\linewidth]{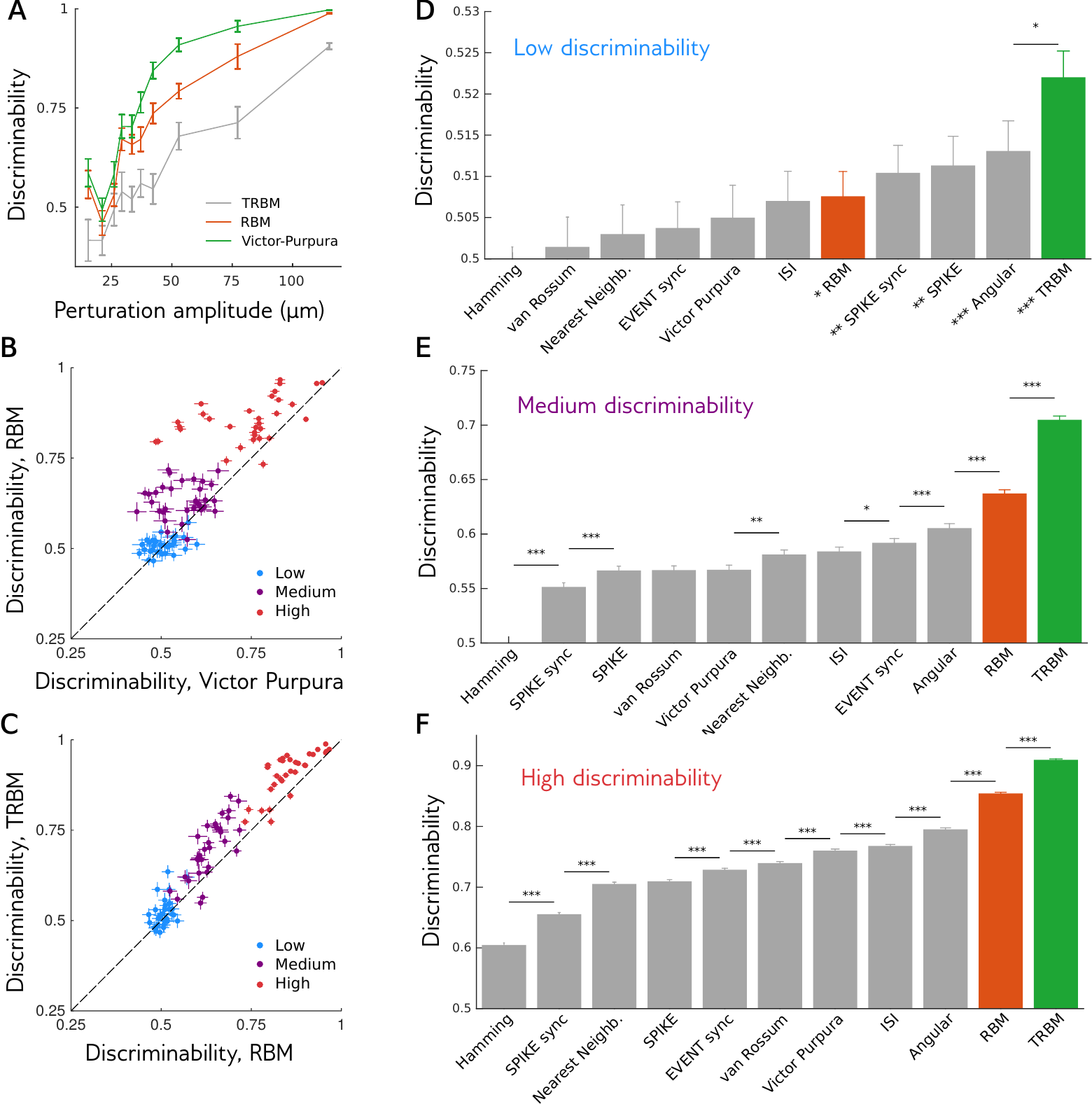}
\caption{
RBM and TRBM metrics outperform classical metrics at discriminating responses.
{\textbf A}, 
Mean discriminability of responses to different amplitudes of an
example perturbation shape, for the optimized Victor Purpura metric or
the RBM and TRBM metrics. Error bars: standard error.
{\textbf B},
Each point represents the mean discriminability for responses with low, medium or high linear discriminability (\cfMM), for one reference trajectory and one perturbation shape, for the Victor Purpura or RBM metric. Error bar: standard error.
{\textbf C}, 
Same as B, but for RBM and TRBM metrics.
{\textbf D}, 
Mean discriminability of responses with low discriminability, across all reference stimuli and perturbation shapes.
Error bars: standard errors.
Stars on top of bars show significant difference in mean discriminability (paired $t$-test, *,**,***: $p$ value lower than 0.05, 0.01 and 0.001).
Stars next to metric names indicate mean discriminability significantly larger than 0.5 ($p<$0.001,unpaired t-test).
\textbf{E,F}, 
Same as D, for responses with medium and high discriminability. All distance had discriminability significantly larger than 0.5.
}
\label{f:discrim__Bar}
\end{figure*}

\subsection*{TRBM metric outperforms other neural metrics at fine discrimination tasks}
We measured the discriminability (Eq.~\ref{eq:discr}) of a perturbation at different amplitudes, using the RBM and TRBM metrics (\cffig\ref{f:discrim__Bar}A, \cfMM).
As expected, the discriminability increased with the perturbation amplitude, with 
small perturbations being hardly discriminable from the reference stimulus (discriminability close to 0.5), and large perturbations almost perfectly discriminable (discriminability close to 1). Since this metric is based on the hidden states, it means that hidden states are informative about the stimulus.
The much better performance of the TRBM, especially for small and medium perturbations, emphasizes the importance of temporal correlations in shaping the metric.
For comparison, we computed the discriminability of the same perturbation for the Victor-Purpura metric \citep{Victor1996} (\cfMM), one of the first  proposed neural metrics which has often been used in the literature to estimate the sensitivity of neural systems \citep{Aronov2003,Chase2006,DiLorenzo2009}.
This metric depends on a  time scale parameter, which we optimized to maximize the mean discriminability of all recorded responses. 
Even with this optimization, the Viktor-Purpura metric discriminated perturbations less well than either the RBM or TRBM metrics, whose parameters were not optimized, across all perturbation amplitudes.

To see if this better performance of our TRBM metrics held for other stimuli, we compared the discrimination capacity of the RBM and TRBM metrics with the Victor-Purpura metric, for 2 different reference stimuli and 16 perturbation shapes for each (Fig.~S1).
For each reference stimulus and perturbation shape, we separated responses in batches of low, medium and high discriminability, 
based on a linear discrimination task independent of any metric (\cfMM).
We computed the mean discriminability of each response batch, for the RBM, TRBM, and Victor-Purpura metrics (\cffig\ref{f:discrim__Bar}B and C).
While responses in the low discriminability batch were poorly separated by all three metrics,
a large majority of responses with medium and high discriminability had larger discriminability for the RBM metric (\cffig\ref{f:discrim__Bar}B), and even larger for the TRBM metric (\cffig\ref{f:discrim__Bar}C), confirming the importance of temporal correlations.

We then compared the RBM and TRBM metrics to other neural metrics from the literature: van Rossum, angular, inter-spike interval (ISI), nearest neighbor, event synchronization, spike synchronization, and SPIKE metrics (definitions in \cfMM), as well as the simple Hamming distance on the binarized responses.
Metrics with free parameters were optimized to maximize their mean discriminability. 
For each metric, we computed the mean discriminability in each batch (low, medium or high discriminability) across all reference stimuli and perturbation shapes (\cffig\ref{f:discrim__Bar}C--E). 
Responses from the low-discriminability batch were hard to distinguish, and only five metrics did significantly better than chance ($p<0.05$ for unpaired \textit{t}-test, \cffig\ref{f:discrim__Bar}C): RBM, spike synchronization, SPIKE, Angular and TRBM metrics. The TRBM metric discriminated responses the best, and was significantly better than the second best, the SPIKE metric ($p=0.014$, paired \textit{t}-test).
For the medium and high discriminability batches, 
the RBM and TRBM metrics greatly outperformed all other metrics. 
Strikingly, in the medium discriminability group, the improvement of discriminability above chance level was {30\% higher for the RBM metric, and 94\% for the TRBM metric, than for the Angular metric}, the most discriminating metric from the literature.

This performance was little affected by the number of hidden units in the RBM and TRBM. The mean discriminability increases and eventually saturates with the number of hidden units (Fig.~S2), indicating that the metric was not sensitive to that precise number, provided that it is large enough. By contrast, the {TRBM}-based metric using the {\em Euclidian} distance between the mean values of the hidden units degraded quickly with the number of units (dashed lines in Fig.~S2). This worse performance may be explained by the fact that some hidden units have little or redundant impact on the activity, but are counted with equal weight in the Euclidian distance. This stresses the importance of accounting for the impact of hidden units on the activity through a distance kernel as in Eq.~\ref{eq:metric}.

Finally, we checked that our conclusions were not affected by the choice of bin size. We repeated all procedures for RBM and TRBM with time bins of size 5, 10 and 40 ms, and obtained  consistent results  (\cfSI).

\setcounter{table}{0}
\renewcommand{\thetable}{S\arabic{table}}%
\setcounter{figure}{0}
\renewcommand{\thefigure}{S\arabic{figure}}%

\section*{Discussion}

We have designed a novel and general method to build a metric from the neural responses of sensory organs, which outperforms all previously defined metrics when trying to discriminate stimuli. This metric called, the TRBM metric, is based on a statistical model of the activity. Importantly, that model is trained in an {\em unsupervised} way, meaning that no knowledge of the stimulus was required in the learning procedure.
Previous work has considered the possibility of constructing `semantic' metrics based on the relationship between the retinal response and the stimuli that evoked them, and proposed a clustering scheme based on that metric \cite{Ganmor2015}. Although such a metric clearly follows the structure of the population code, as does our own, it presents an extreme case of supervised learning, 
as building this metric in practice requires to know the probability of response triggered by all possible stimuli, which the brain cannot do in practice.
By contrast, the TRBM metric discriminates stimuli well, 
but rather than being learned from the stimulus-response dictionary, it naturally emerges from learning the structure of the retinal code, which can be done with a reasonable amount of data and without any information about the stimulus. This way of constructing a metric suggests a realistic strategy for the brain to learn to discriminate stimuli. 
Note that many neural metrics require to tune parameter values to maximize performance---a supervised process as it uses knowledge of the stimulus in evaluating discrimination {\em a posteriori}. 
Strikingly, even after this optimization, the TRBM metric outperforms all metrics we found in the literature.

Although we have motivated introducing the TRBM for defining a metric, statistical models of population activity deserve attention in their own right. In this regard, the TRBM provides an alternative to existing approaches that is both accurate and tractable. The spiking responses of retinal ganglion cells at a given time are strongly correlated \cite{Schneidman2003,Schneidman2006}, and various strategies have been proposed to model their collective, synchronous (same time-bin) activity. Central to this effort are models based on the principle of maximum entropy  \cite{Schneidman2006,Shlens2006,Tang2008,Shlens2009}, which allow for an explicit mapping onto models of Ising spins from statistical mechanics, also known as Boltzmann machines \cite{Ackley1985}. However these models are often hard to learn in practice, and they need additional terms \cite{Tkacik2014searching,Gardella2016} or non-linearities \cite{Humplik2017} to explain higher-order statistics such the distribution of total number of spikes. It also is unclear how to exploit their structure to derive a metric.

As an alternative to Ising models, RBMs were applied to the correlated activity in cortical micro-columns \cite{Koster2014} and in the retina \cite{Schwab2013,Humplik2016}. Our results confirm their ability to describe the synchronous collective activity in the retina, including pairwise correlations and the distribution of total numbers of spikes. Previous work also showed that the hidden units of a variant of the RBM conveyed information about the stimulus, although this was only made possible by the small number of used stimuli \cite{Zanotto2017}.
All these models ignore correlations between spikes in different time bins, which play an important role as we have shown here. Maximum entropy models were generalized to account for correlations across time, but they were either practically intractable for large populations \cite{Vasquez2012,Nasser2013}, or only focused on the total number of spikes \cite{Mora2015}. The TRBM can reproduce pairwise and higher-order correlations with high accuracy across different time bins, and this with a reasonable amount of parameters and with relative computational ease. We therefore expect the TRBM to be useful in describing the stimulus-independent activity of neural populations in a variety of contexts \cite{Yu2008,Kampa2011,Bathellier2012,Truccolo2014}.

Also using latent variables, a Hidden Markov Model (HMM) was proposed to describe retinal activity, in which the response is controlled by a hidden categorical variable following a Markov chain \cite{Prentice2016}. However, to reflect the diversity of responses, the number of categories encoded by the hidden variable should grow exponentially with the population size, which is impractical computationally and biologically. By contrast, the number of configurations of the hidden state in the TRBM grows exponentially with the number of hidden units, making easy its application to large populations. Continuous latent variables have also been proposed to account for neural correlations in cortical networks, {such as in} the linear dynamical system \citep{Gao2016}. It could be interesting to apply such models in the retina. 
However, complex computational techniques are needed to infer them, while the TRBM is relatively simple to learn.

Our work presents an example of unsupervised learning (inferring the TRBM) which proves predictive in a supervised task (stimulus discrimination). This is reminiscent of the technique of {unsupervised} {`pre-training'} \citep{Hinton2006}, which is used in machine learning when only a few examples are available to improve model performance.
The link to machine learning suggests to consider ``deep'' extensions of the RBM, with several layers of hidden variables \citep{Salakhutdinov2009deep}, from which more general metrics could be derived. It has been shown that deep (artificial) neural networks achieve higher discrimination power than RBMs when dealing with complex stimuli such as natural scenes, and such complex architectures could lead to better metrics in our case as well.

The TRBM was trained on responses to the random motion of a bar. 
One may wonder how well the TRBM and its derived metric would perform when confronted with stimuli from different statistics (e.g. random checkerboard, natural movies). One possibility is that the brain constantly re-learns the metric depending on the visual stimulus, or alternatively learns a universal metric that performs well across a wide random of stimulus conditions. 
To address this question, we would need not only to display several different stimulus ensembles, but also to design an experimental procedure for comparing responses to pairs of close enough stimuli in that ensemble. 
Studying this question is an interesting avenue for future research.

While neural metrics may not be explicitly estimated by the brain, our TRBM metrics have a natural biological implementation that suggests how a downstream population could discriminate responses to different stimuli. 
Hidden units could be implemented by a population of downstream neurons, with a simple response function: a weighted sum followed by a nonlinearity (\cfeq\ref{eq:hprobaTRBM}).
This is reminiscent of a neuron summing responses from upstream cells, weighted by synapses' strengths, with delays to account for time lags.
Indeed, it was shown that networks of spiking neurons can learn their synaptic weights to approximate Restricted Boltzmann Machines \cite{Nakano2015}. 
One can simplify our TRBM metric by linearizing the dependence of the hidden units as a function of activity (see \cfMM). Doing so leads to a metric that readily generalizes to continuous times, where the binning of time disappears. The metric is then simply given by a sum over pairs of spikes, with coefficients depending on the identity of the spiking neurons and the delay between them. We showed that this simplified ``continuous'' TRBM performs almost as well as the full TRBM metric (Fig.~S3). The continuous TRBM metric could be implemented by simple summation of spikes with time delays. 

In summary, the TRBM provides insights into biologically possible representations of the stimulus with high discrimination capabilities, without the need for any supervised training. None of the properties of the TRBM and its derived metric are expected to be specific to the retina, and our method could be readily applied to other sensory neural circuits.

\begin{figure}
\begin{center}
\includegraphics[width=1\linewidth]{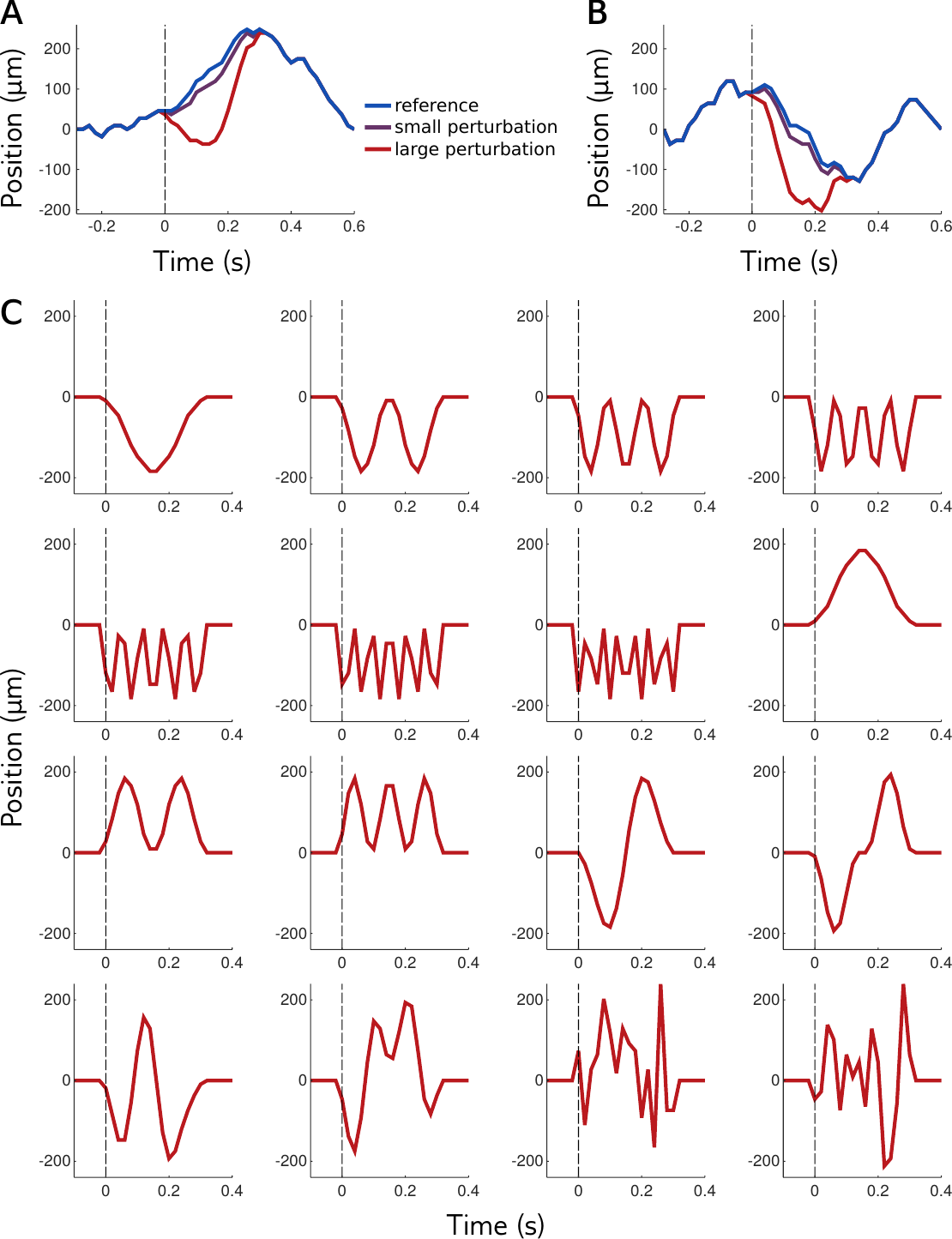}
\end{center}
\caption{
Temporal trajectory of the bar for the (A) first and (B) second
reference stimuli. (C) Perturbations of 16 different ``shapes'' (in
the space of bar trajectories), all
represented here, are added to either of the two reference stimuli,
with varying amplitudes.
}
\label{f:shapes}
\end{figure}

\begin{figure}
\begin{center}
\includegraphics[width=1\linewidth]{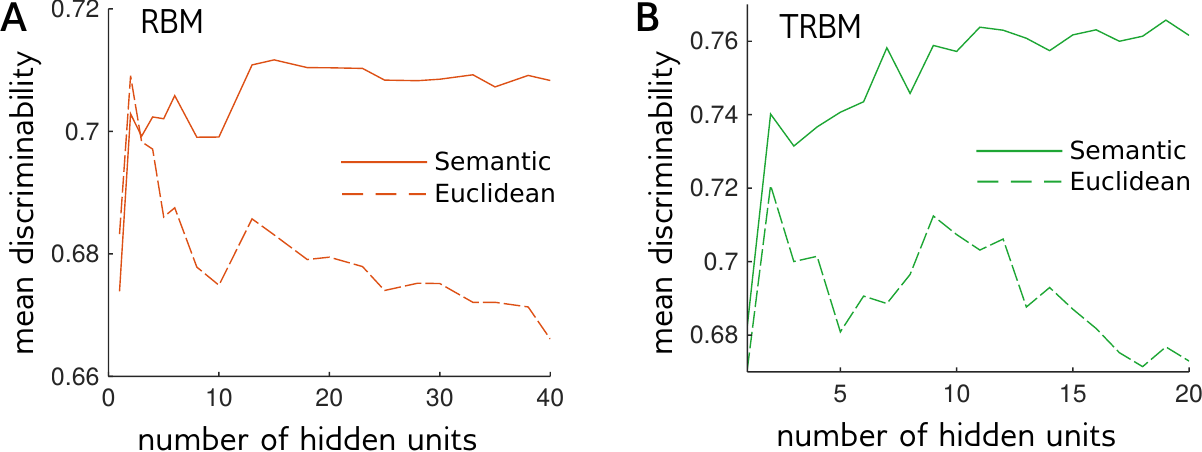}
\end{center}
\caption{
RBM and TRBM performances are not affected by large numbers of parameters.
{\textbf A}, 
Mean discriminability of responses to perturbations, measured by the RBM metric for different numbers of hidden units. The discriminability for the \textit{semantic} RBM metric (used by default in this paper, \cfMM) increases with the number of hidden units, reaches a maximum and decays. 
On the contrary, for the simpler Euclidean RBM metric, the discriminability of responses decreases with the number of hidden units.
\textbf{B},
Same as A, for a TRBM with maximum delay of 80 ms between neurons and hidden units.
}
\label{f:fit_vs_discrim}
\end{figure}

\begin{figure}
\begin{center}
\includegraphics[width=1\linewidth]{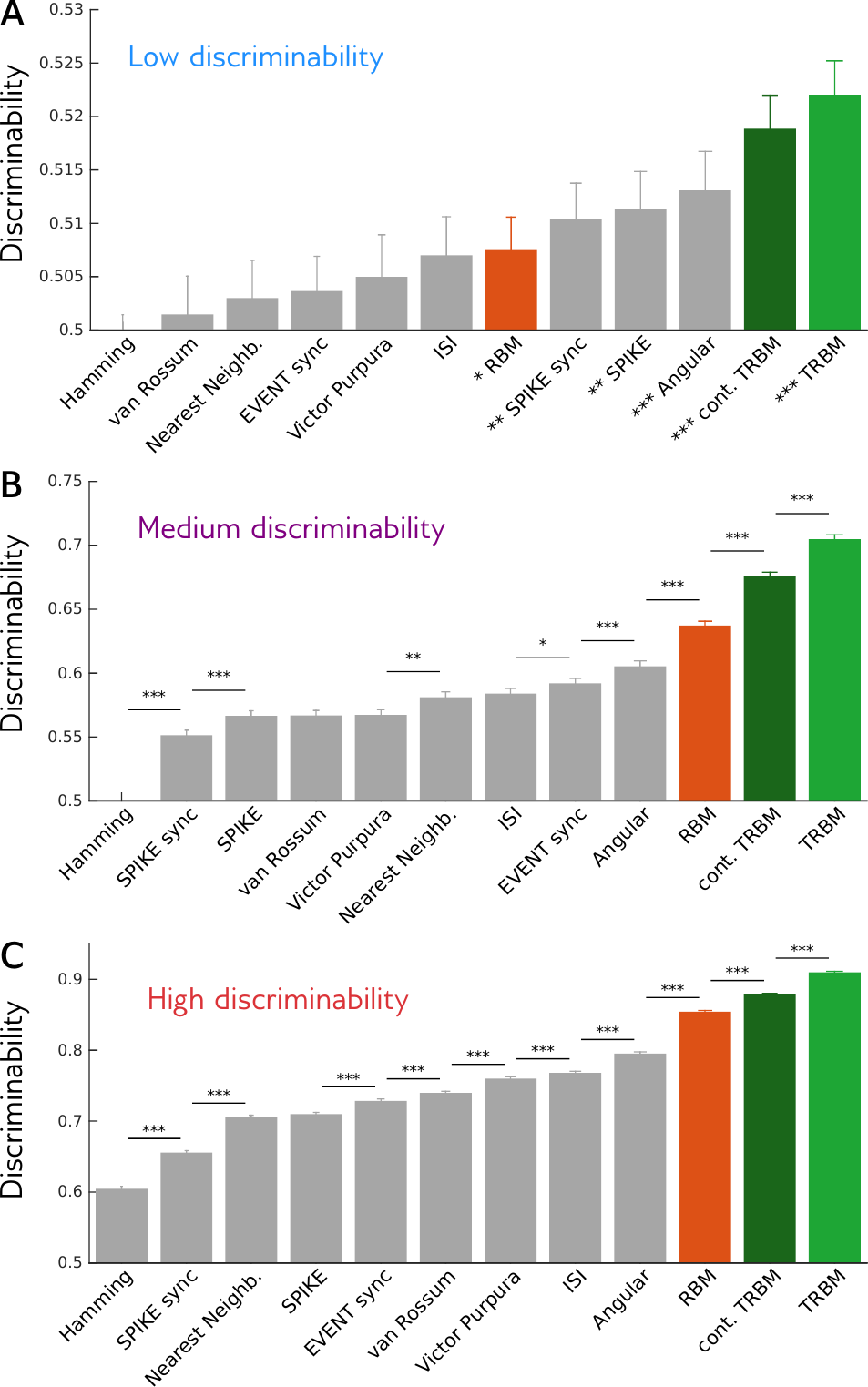}
\end{center}
\caption{
Same as Figure 5D-F, but with the continuous TRBM added (see Materials
and Methods for details).
}
\label{f:shapes}
\end{figure}

\section{\cfMM}

\subsection{Electrophysiology}
We analyzed previously published \textit{ex vivo} recordings of retinal ganglion cells from a male Long Evans rat \citep{Ferrari2016closed}. In brief, the animal was killed according to institutional animal care standards. The retina was extracted from the animal, maintained in an oxygenated Ames solution, and recorded on the ganglion cell side with a 252-electrode array. Spike sorting was performed with custom software \citep{Marre2012}, and $N=60$ neurons were selected for the stability of their spike waveforms and firing rates, the lack of refractory period violation, and the consistency of their responses to repeated stimuli.

\subsection{Stimulus}
The stimulus was a movie of a white bar on a dark background projected at 50 Hz with a digital micromirror device. The bar had intensity $7.6\times10^{11}$~photons.cm$^{-2}$.s$^{-1}$, and 115~\textmu m width.

The bar trajectory was composed of two interleaved parts.
{The first part was composed of multiple 0.9~s-long snipets of random motion (a Brownian motion with a restoring force).}
The second part was composed of 391 repetitions of two trajectories lasting 0.9~s each, which we call reference trajectories (Fig.~S1 A-B). Both parts were generated with the same random motion. The two parts were alternated for a total of 15331~s.

We also presented perturbations of the reference trajectories. Perturbations were small changes affecting each reference trajectory in its middle portion, between 280 and 600~ms. 
Perturbations varied both in shape and in amplitude: we used 16 different perturbation shapes (Fig.~S1 C), each presented at different amplitudes (\cffig\ref{f:closed_loop}B). 
The amplitude of perturbations was adapted online: large enough so they could be   discriminated from reference trajectories, but small enough so they would not be discriminated perfectly with any metric. 

Formally, if we denote a perturbation by its discretized time series with time step $\Delta_t=20$ ms, $S=(S_1,\ldots,S_{16})$, it can be decomposed as $S=A\times P$, where $A^2={(1/16)\sum_{k=1}^{16} S_k^2}$ is the amplitude, and $P=S/A$ the shape. More details can be found in \cite{Ferrari2016closed}.

For each perturbation, the response was considered from 160 ms after the start of the perturbation until 140~ms after its end, so that responses lasted 300 ms (\cffig\ref{f:closed_loop}C).

\subsection{Models of response distributions}
In order to model neural correlations, the response of $N$ neurons was binarized into time bins of size $\Delta_t = 20$~ms: $\s_{i} = 1$ if neuron $i$ spiked, and 0 otherwise (Fig.~\ref{f:setup}). 

\subsubsection{Restricted Boltzmann Machine}
We used Restricted Boltzmann Machines (RBM) to model the probability of responses within a single time bin. RBM are statistical models with no direct interactions between the $N$ neurons. 
Instead, the neurons interact with $M$ binary latent variables, termed hidden units.
There is no direct interactions between hidden units either. 
If we call $h_{j}$ the hidden unit $j$, the joint probability between neurons and hidden units takes the form $P(\bm{\s},\bm{h}) \propto e^{-E}$ with:
\eq{
-E\! = \! \sum_{i=1}^N a_i \s_{i} + \sum_{j=1}^M b_j h_{j} + \sum_{i,j=1}^{N,M} W_{ji} \s_{i} h_{j} \;,
}
or in matrix form: 
\eq{
-E\! = \bm{a \s} + \bm{b h} + \bm{h}^\intercal \bm{W \s}\;,
\label{eq:E_RBM}}
where the bold font stands for vectors and matrices, and $\bm{h}^\intercal$ stands for the transpose of $\bm{h}$. The hidden variables do not necessarily correspond to existing entities interacting with neurons. Instead, they are effective variables that are used to capture correlations between neurons.

This model is convenient to simulate because neurons (resp. hidden units) conditioned by hidden units (resp. neurons) are independent. 
Namely, given a state of hidden units $\bm{h}$, neurons are independent and:
\eq{
P(\s_i=1|\bm{h}) = f\left( a_i + \sum_j W_{ji} h_j \right)
\label{eq:sprobaRBM}
}
where $f(x) = 1/(1+e^{-x})$ is the sigmoid function.  The conditional probability of hidden units given the neurons can be computed with a similar formula.

We learned the RBM on $80\%$ of responses to the random trajectories of the bar, with $M = 20$ hidden units.
We inferred the model by maximizing the likelihood using persistent contrastive divergence \citep{Tieleman2008} with 200 epochs and minibatches of size 10.
We used the momentum method \citep{Fischer2012}, which is known to accelerate learning, with a momentum coefficient of 0.9.
This method updates parameters in a direction proportional to the sum of the likelihood gradient and the parameter update at the previous step. 
For regularization, we used a weight decay parameter of $10^{-5}$ \citep{Krizhevsky2012}, meaning that the  objective function maximized during learning was the sum of the log-likelihood and of the Euclidean norm of coupling parameters $\bm{W}$, weighted by a factor $10^{-5}$.
For computation of model statistics, we simulated the model using block Gibbs Sampling \citep{Fischer2012}. Final model statistics (Fig.~\ref{f:stats_RBM}) were computed with simulations with 300 steps on responses spanning $2\times10^6$ time bins.   

\subsubsection{Temporal Restricted Boltzmann Machine}
In order to model interactions between neurons across multiple time bins, we use a more complex RBM, allowing connections between neurons and hidden units in different time bins.
We call it the Temporal Restricted Boltzmann Machine (TRBM). 
In order to limit the number of parameters, this model is convolutional \citep{Lee2011}, meaning that the interaction between a neuron and a hidden unit does not depend on their absolute time, but only on the delay $d$ between them. 
We define $\s_{ik}$ as the response of neuron $i$ in time bin $k$.
The probability of neurons and hidden units during $K$ time bins takes the form $P(\bm{\s},\bm{h}) \propto e^{-E}$, with:
\eq{
- \!E\! = \sum_{k=1}^K \! \left[\! \sum_{i} a_i \s_{ik} + \sum_{j} b_j h_{jk} + \sum_{ij} \! \sum_{d=0}^{D-1} W_{dji}   \s_{ik} h_{j,k+d} \!\right] \!,
}
where the time span of interactions between hidden and visible units is $D$. 
It can be noted that in the case $D=1$, the TRBM consists of an independent RBM in each time bin. The TRBM can also be written in matrix form:
\eq{
- \!E\! = \sum_{k=1}^K  \bm{a} \bm{\s}_{k} +  \bm{b} \bm{h}_{k} + \sum_{d=0}^{D-1} \bm{h}_{k+d}^\intercal \bm{W}_{d} \bm{\s}_{k} \;,
\label{eq:E_TRBM}}
where $\bm{W}_d$ are matrices of size $M \times N$. 

The parameters can be learned independently of the length of responses considered, and can be used to model responses of different lengths.
Visible (resp. hidden) units at the time boundaries depend on hidden (resp. visible) units outside boundaries.
However, given a finite-size response we can still compute exactly the probability of some hidden units using:
\eq{
P(h_{jk}=1|\bm{\s}) = f\left( b_j + \sum_{d=0}^{D-1}\sum_i W_{dji} \s_{i,k-d}  \right).
\label{eq:hprobaTRBM}}
{Therefore, during learning we only consider hidden units in time bin $D-1$ and later.}

We learned the TRBM on $80\%$ of responses to random bar trajectories, with $M = 10$ hidden units and $D=5$. We inferred the model using persistent contrastive divergence with 400 epochs, minibatches of 2 responses of length 820~ms ($K=41$ bins), momentum coefficient 0.9, and weight decay of $10^{-5}$.

For computation of model statistics, we simulated the model using block Gibbs sampling. In order to avoid dependence on times out of boundaries, we used cyclic boundary conditions. Final model statistics (Fig.~\ref{f:stats_TRBM}) were computed with simulations with 300 steps on responses spanning $2\times10^6$ time bins.   

A slightly different model was proposed  \cite{Sutskever2007learning} and later simplified \cite{Sutskever2008recurrent}, which was also called temporal restricted Boltzmann machine. Contrary to the model presented here, their model also allows for neuron-to-neuron and hidden unit-to-hidden unit interactions across different times. Furthermore, the dependence on the past takes has a slightly different form. At each time, the joint probability of neurons and hidden units is an RBM, with field vectors $\bm{a}$ and $\bm{b}$ depending on past values of neurons and hidden units. This difference in form has consequences for model computations. For example, it is not possible to compute exactly the probability of hidden states given neural responses (\cfeq\ref{eq:hprobaTRBM}), which is used for learning.

\subsubsection{Response statistics}
We measured how well models of population response could predict multiple statistics that are classically encountered in the literature.

\textit{Firing rate}, or spiking frequency \citep{Schneidman2006,Tang2008}. 
It is equal to $\< \s_{ik} \>_k/\Delta_t$ for neuron $i$, where $\Delta_t$ is the length of time bins. Here $\<.\>_k$ stands for the mean across time bins $k$.

\textit{Pairwise correlation} \citep{Schneidman2006,Tkacik2014searching}: Pearson correlation between neurons $i$ and $i'$: $\text{corr}(\s_i,\s_{i'})=\text{cov}(\s_i,\s_{i'})/\sqrt{\text{var}(\s_i)\text{var}(\s_{i'})}$, where the covariance is $\text{cov}(X,Y) = \<X,Y\> -  \<X\>  \<Y\>$ and the variance is $\text{var}(X) =\text{cov}(X,X)$.

\textit{Cross-correlation}, the correlation between neuron $i$ and $i'$ with a delay $d$ between them, $\rho_{ii'd} = \text{corr}_k(\s_{ik},\s_{i',k+d})$, , where $\text{corr}_{k}$ is the correlation across time bins $k$.

In order to quantify how well a model predicted cross-correlations, we computed the explained variance of cross-correlations.
We first computed the variance of cross-correlations for different time delays $d$, $ \text{var}_{i,i'}( \rho_{\text{data},ii'd} )$, where $\text{var}_{i,i'}$ is the variance across all pairs of neurons.
 $\rho_\text{data}$ was computed on a testing set not used for training. 
 We then computed the variance explained by the model:
\eq{
\text{e.v.} = \text{var}_{i,i'}( \rho_{\text{data},ii'd} ) - \< \left(\rho_{\text{data},ii'd}  - \rho_{\text{model},ii'd} \right)^2 \>_{ii'}
}  
where $\rho_\text{model}$ is the cross-correlation predicted by the model.

On \cffig\ref{f:stats_RBM}E, we show the fraction of explained variance, equal to the ratio between the explained variance and the variance, for cross-correlations at a single time delay.

On \cffig\ref{f:stats_TRBM}F,G, we compute the fraction of explained variance across 8 time bins. Here we grouped all cross-correlations with multiple time delays for computation of the variance and explained variance. 

\textit{Population count}: number of spikes in the population during $L$ time bins. At time $k$ it is $ \sum_{l =0}^{L-1} \sum_{i=1}^N \s_{i,k+l}$.

\subsubsection{Likelihood}
We computed the mean log-likelihood of responses for the RBM and TRBM models:
\eq{
\mathcal{L} = \< \log P_\text{model}(\bm{\s}) \>_ {\bm{\s}} \;,
}
where the empirical mean is computed across responses in the training or testing sets. For the RBM, we computed the likelihood for responses of length 20~ms (single time bin).
For the TRBM, we computed the likelihood for responses of length 100~ms (5 time bins). 

The probability of response $\bm{\s}$ for the RBM, marginalized across all possible hidden states $\bm{h}$, has the form $P(\bm{\s}) = (1/Z)  e^{-E(\bm{\s})}$, where $Z$ is a normalization constant called partition function, and:
\eq{
-E(\bm{\s}) = \bm{a \s} + \sum_j \log\left( 1 + e^{b_j +  \bm{W}_j \bm{\s} }\right)  \;,
}
where $\bm{W}_j$ is the $j^{th}$ row of $\bm{W}$ \citep{Fischer2012}.

A similar equation exists for the TRBM. Due to temporal correlations, the probability of responses across time bins $k=1,...,K$ also depends on responses in time bins $k<1$ and $k>K$. In order to compute exactly the distribution of responses in time bins $k=1,...,K$, one needs to marginalize over all possible responses in other time bins. As this is intractable, we approximated the probability of response $\bm{\s}$ in time bins $k=1,...,K$ using:
\eq{
\begin{split}
&- E(\bm{\s}) \approx \sum_{k=1}^K  \bm{a} \bm{\s}_{k} \\ & \ +  \sum_{b=1}^{K+D-1} \sum_j  \log \left( 1 + \exp\left( b_j + \sum_{d=0}^{D-1} \bm{W}_{dj} \bm{\s}_{k-d} \right) \right),
\end{split}
\label{eq:TRBMlike}
}
where $\bm{W}_{dj}$ is the $j^{th}$ row of matrix $\bm{W}_{d}$. In \cfeq\eqref{eq:TRBMlike}, we replaced the response $\bm{\s}_k$ outside time bins $k=1...K$ by the mean response $\< \bm{\s} \>$ predicted by the model.

We computed the normalizing constant $Z$ for the RBM and TRBM using Annealed Importance Sampling \citep{Neal2001annealed,Salakhutdinov2008learning} with 5000 intermediate temperatures and 5000 responses generated at each temperature.

\subsection{Neural metrics}
The response of a population of neurons consists in a series of action potentials, or spike train. We note $R = (t_{in})_{in}$ the population response, with $t_{in}$ the time of the $n^{th}$ spike from neuron $i$. 
Neural metrics are functions that associate a non negative value to each pair of responses $R^{(1)}$ and $R^{(2)}$ (exponents with parenthesis are indices). As such, they are a measure of the dissimilarity between responses.
In the following we present multiple neural metrics that can be found in the literature, and then introduce new metrics based on the RBM and TRBM. 
When a metric from the literature was only defined for single neurons, we adapt it to a population by summing the metric for each neuron.

The first three metrics are functional metrics \citep{Paiva2010}: responses are first mapped onto time dependent vectors, and the metric is defined in this functional space.
The remaining metrics are defined directly on spike trains. 

\subsubsection{van Rossum metric}
The van Rossum metric is a kernel-based metric. To map a response $R$ to a time dependent vector $\bm{v}$, each neuron's spike train is convolved with a kernel $H$: $v_i(t) = \sum_n H (t-t_{in})$. We then take the Euclidean distance between convolved spike trains. 
\begin{equation}
d_{\text{van Rossum}}(R^{(1)},R^{(2)})^2  = \sum_{i} \int | v_i^{(1)}(t) - v_i^{(2)}(t)|^2 \, \text{d}t
\end{equation}
Classically $H$ is a decaying exponential: $H(t) = e^{-t/c}$ if $t\geq 0$, $0$ otherwise \citep{VanRossum2001} with $c$ a time constant. We optimized $c$ to maximize mean response discriminability across all responses to perturbations (\cfSI, time constants for all metrics were optimized in the same way), and found $c= 630$~ms here. This constant might seem large compared to time constants of metrics presented below, but they are actually not directly comparable, as they are not on the same scale. This is due to the asymmetry of $H$: the van Rossum metric still takes spike times into account in the limit of $c$ infinitely large. Indeed, for large $c$, $\bm{v}(t)$ is proportional to the number of spikes that happened before $t$. On the opposite, the metrics presented below which depend on a time constant only compare the total number of spike for each neuron when their time constant is large, with no information about their timing.

A Gaussian kernel is sometimes also considered: $H(t)=e^{-t^2/2 c^2}$ \citep{Houghton2011}. Even after optimizing the time scale, it always discriminated less well than the exponential kernel. It is therefore not shown here.

\subsubsection{Angular metric}
The angular metric uses the same vector mapping as the van Rossum metric, but measures the angle between corresponding vectors \citep{Schreiber2003}: 
\begin{equation}
d_{\text{angular}}(R^{(1)},R^{(2)})^2  = \sum_i \; \arccos \dfrac{ \< v_i^{(1)} , v_i^{(2)} \> }{ | v_i^{(1)} |_2 \; | v_i^{(2)} |_2}
\end{equation}
where $\< . , . \>$ and $|.|_2$ are the scalar product and Euclidean norm respectively: $\< x , y \> = \int x(t) \; y(t) \, \text{d}t $, $|x|_2^2 = \< x, x\>$. 

In order to account for responses with no spike, we add an offset $\alpha$ to convolved spike trains: $v_i(t) = \sum_n H (t-t_{in}) + \alpha$. 
We used a Gaussian kernel, optimized $\alpha$ and the time constant $c$ (\cfSI) and found  $\alpha = 10^{-5}$ (for kernel with integral norm 1~s) and $c = 80$~ms.

\subsubsection{Inter-Spike Interval metric}
The Inter-Spike Interval (ISI) metric measures the dissimilarity between responses inter-spike interval profiles $\bm{\nu}$ \citep{Kreuz2007,Mulansky2015}. For each neuron $i$ and time $t$, we define $\nu_i(t) = t_{i,n+1} - t_{in}$, where $t_{in}$ (resp. $t_{i,n+1}$) is the first spike before (resp. after) $t$ for neuron $i$. The ISI metric is then:
\begin{equation}
d_{\text{ISI}}(R^{(1)},R^{(2)})  = \sum_{i} \int \dfrac{|\nu_i^{(1)}(t) - \nu_i^{(2)}(t)|}{\max\!\left[\nu_i^{(1)}(t), \nu_i^{(2)}(t)\right]} \, \text{d}t
\end{equation}
We used an edge-correction in order to estimate $\bm{\nu}$ before the first spike and after the last \citep{Mulansky2015}. The ISI metric has no parameter.

\subsubsection{Victor-Purpura metric}
The Victor-Purpura metric \citep{Victor1996} is an edit-length metric: the distance between two spike trains is the minimal cost necessary to transform a spike train into the other. Deleting or adding a spike costs +1, whereas moving a spike by $\Delta t$ has a linear cost $q \Delta t$. We optimized $q$ to maximize mean response discriminability (\cfSI) and found $q=13$~s$^{-1}$.

\subsubsection{Nearest-Neighbor metric}
The Nearest-Neighbor metric measures the similarity in spike times \citep{Hunter2003}. 
Given two population responses $R^{(1)} = (t^{(1)}_{in})_{in}$ and $R^{(2)} = (t^{(2)}_{in'})_{in'}$, we compute the distance between them by computing, for each spike $n$ from neuron $i$, the time difference with the nearest spike in the other response: $\Delta^{(1)}_{in} = \min_{n'} |t^{(1)}_{in} - t^{(2)}_{in'}|$, and symmetrically for $\Delta^{(2)}$. The distance between the two population responses is then:
\begin{equation}
d_{\text{NN}}( R^{(1)}, R^{(2)} ) = \sum_i 2 - \< \exp( - \dfrac{\Delta^{(1)}_{in}}{c}) \>_n - \< \exp( - \dfrac{\Delta^{(2)}_{in'}}{c}) \>_{n'}
\end{equation}
where $\< . \>_n$ is the mean across spikes. We optimized $c$ (\cfSI) and found $c = 50$~ms.

\subsubsection{Event and Spike Synchronization metrics}
The synchronization metrics are based on an instantaneous coincidence detector $F$ \citep{Quiroga2002,Kreuz2015spiky,Mulansky2015}. For each spike of $R^{(1)}$, $F^{(1)}_{in}$ is equal to 1 if there is a coinciding spike in $R^{(2)}$, and 0 otherwise:
\begin{equation}
F^{(1)}_{in} = \begin{cases}
1 &\text{ if }\ \min_{n'} |t^{(1)}_{in} - t^{(2)}_{in'}| < \tau_{in}  \\  0 &\text{else}
\end{cases}
\end{equation}
We compute $F^{(2)}$ symmetrically. The synchronization metrics are then:
\begin{equation}
d_{\text{Sync}}( R^{(1)}, R^{(2)} ) = \sum_i \left( 1 - \< F_{in} \>_n \right)
\end{equation}
where the average is across all spikes in $F^{(1)}$ and $F^{(2)}$. 

For the Event synchronization metric, the time scale is fixed: $\tau_{in}=c$. We optimized $c$ (\cfSI) and found $c= 50$~ms.

For the Spike Synchronization metric, the time scale is automatically adapted to the local firing rate of the responses, so it has no parameter. For a spike $t^{(1)}_{in}$ with closest spike in the other response $t^{(2)}_{in'}$, we take:
\eq{
\tau_{in}  = \dfrac{1}{2} \min( & t^{(1)}_{i,n+1}\!-\! t^{(1)}_{in} \;, \; t^{(1)}_{in} \!-\! t^{(1)}_{i,n-1} \;, \\  & t^{(2)}_{i,n'+1} \!-\! t^{(2)}_{in'} \;, \; t^{(2)}_{in'} \!-\! t^{(2)}_{i,n'-1}).
}

\subsubsection{SPIKE metric}
The SPIKE metric is based on the SPIKE dissimilarity profile $S(t)$, measuring differences in timing of spike events \citep{Kreuz2011,Kreuz2013,Mulansky2015}. 
For neuron $i$ and time $t$ between spike times $(t^{(1)}_{in} , t^{(1)}_{i,n+1})$ in response $R^{(1)}$, we set $\zeta^{(1)}_i$ a weighted average between times to closest spikes in the other response, $\Delta^{(1)}_{in}$ and $\Delta^{(1)}_{i,n+1}$, defined in the Nearest-Neighbor metric:
\eq{
\zeta^{(1)}_i(t) = \dfrac{ \left(t^{(1)}_{i,n+1} - t\right) \Delta^{(1)}_{in} + \left(t - t^{(1)}_{in}\right) \Delta^{(1)}_{i,n+1}}{t^{(1)}_{i,n+1} - t^{(1)}_{in}}
}
We call $\zeta^{(2)}$ the corresponding average for response $R^{(2)}$. $S$ is then a weighted sum between $\zeta^{(1)}$ and $\zeta^{(2)}$:
\eq{
S_i = \dfrac{\zeta^{(1)}_i \nu^{(2)}_i + \zeta^{(2)}_i \nu^{(1)}_i}{\tfrac{1}{2}(\nu^{(1)}_i + \nu^{(2)}_i)^2}
}
with $\bm{\nu}$ the previously defined inter-spike interval profile. The SPIKE metric is:
\eq{
d_{\text{SPIKE}}(R^1, R^2) = \sum_i \int S_i(t) \;\text{d}t
}
The SPIKE metric has no parameter.

\subsubsection{RBM metric}
We first define metrics for responses in a single time bin, and then generalize to longer responses.
We designed RBM metrics, such that the distance between binned responses $\bm{\s}^{(1)}$ and $\bm{\s}^{(2)}$ depends on the difference between the probabilities of hidden units conditioned by neural responses,  $P(\bm{h}|\bm{\s}^{(1)})$ and $P(\bm{h}|\bm{\s}^{(2)})$. 

There are multiple ways to compute a difference between distributions, such as the Kullback-Leibler divergence, but we aim at finding a metric that is convenient for computation. We notice that hidden units are binary and independent when conditioned by a neural response, so $P(\bm{h}|\bm{\s}^{(u)})$ for $u=1,2$ is fully characterized by its mean $\< \bm{h} | \bm{\s}^{(u)} \>$. Therefore we chose to measure the difference between the probabilities $P(\bm{h}|\bm{\s}^{(u)})$ as a difference between mean hidden states: $\bm{\Delta h} = \< \bm{h} | \bm{\s}^{(1)} \> - \< \bm{h} | \bm{\s}^{(2)} \>$.

The difference between those vectors was measured in two different ways, an Euclidean and a \textit{semantic} metric, simply refered to as RBM metric in the main text. 
The Euclidean metric for the RBM is
\eq{
d_\text{Eucl. RBM}(\bm{\s}^{(1)},\bm{\s}^{(2)}) = \vert|\;\bm{\Delta h}\;\vert|_2 .
}
For the semantic RBM metric, we reasoned that two states of hidden units are similar if they trigger similar neural responses. We first define a metric between states of hidden units (termed hidden states) that takes this consideration into account, and then apply it to measure the difference between mean hidden states $\bm{\Delta h}$.

We define a metric in the space of hidden states, such that the distance between hidden states  $\bm{h}^{(1)}$ and $\bm{h}^{(2)}$ depends on the difference between the probability of neural responses they trigger, $P(\bm{\s}|\bm{h}^{(1)})$ and $P(\bm{\s}|\bm{h}^{(2)})$.
It can be shown from \cfeq\eqref{eq:E_RBM} that if the difference between $\bm{h}^{(1)\intercal} \bm{W} \bm{\s}$ and  $\bm{h}^{(2)\intercal} \bm{W} \bm{\s}$ is always constant across different values of $\bm{\s}$, then $P(\bm{\s}|\bm{h}^{(1)})$ and $P(\bm{\s}|\bm{h}^{(2)})$ are equal. 
Thus if $\left( \bm{h}^{(1)} - \bm{h}^{(2)}\right)^\intercal \bm{W \s}$ has only small fluctuations when $\bm{\s}$ is generated by the RBM model, then $\bm{h}^{(1)}$ and $\bm{h}^{(2)}$ have a similar influence on neural responses. 
They can have very different probabilities to happen, but when they do they co-occur with similar neural responses.
We thus chose to measure the distance between $\bm{h}^{(1)}$ and $\bm{h}^{(2)}$ as $\text{var}_\sigma \!  \left[(\bm{h}^{(1)}-\bm{h}^{(2)})^\intercal \bm{W} \bm{\s}\right]$, where var$_\sigma$ is the variance across neural responses predicted by the model. If this value is 0, then $P(\bm{\s}|\bm{h}^{(1)})$ and $P(\bm{\s}|\bm{h}^{(2)})$ are the same. 
 The semantic RBM metric between 2 responses is then:
\eq{
d_\text{RBM}(\bm{\s}^{(1)},\bm{\s}^{(2)})^2 &= \text{var}_\sigma\!\left[\; \bm{\Delta h}^\intercal \bm{W \s} \; \right]\\
&= \bm{\Delta h}^\intercal \bm{W C W}^\intercal \bm{\Delta h} 
}
where $\bm{C}$ is the covariance matrix of neural responses predicted by the model.

As we show in \cffig\ref{f:fit_vs_discrim}, the semantic metric is better at discriminating responses than the Euclidean one, as it is less affected by the redundancy between the parameters of the RBM. 
In this article we always take the semantic metric by default, unless explicitly stated.

Finally, the RBM metric between responses lasting multiple time bins is:
\eq{
d_{\text{RBM}}(R^{(1)},R^{(2)})^2 = \sum_k  d_{\text{RBM}}(\bm{\s}^{(1)}_k, \bm{\s}^{(2)}_k)^2.
}

\subsubsection{TRBM metric}
The RBM is a special case of the TRBM where neurons are only connected to hidden units in the same time bin. We thus define TRBM metrics so that they are consistent with the RBM metrics. We define the vector $\bm{\Delta h} = \< \bm{h} | \bm{\s}^{(1)} \> - \< \bm{h} | \bm{\s}^{(2)} \>$, with indices $(j,k)$ over hidden units $j$ and time bins $k$. The TRBM metrics are then:
\eq{
d_\text{Eucl. TRBM}(R^{(1)},R^{(2)})^2  = \sum_k \vert| \bm{\Delta h}_{k} \vert|_2^2 \, .
}

\eq{
d_{\text{TRBM}} (R^{(1)},R^{(2)})^2 =  \text{var}_\sigma\!\left[ \sum_k  \sum_{d=0}^{D-1} \bm{\Delta h}_{k+d}^\intercal \bm{W}_{d} \bm{\s}_{k}  \right]\,.
\label{eq:distTRBM}
}

The later can also be written in matrix form:
\eq{
d(\bm{\s}^{(1)},\bm{\s}^{(2)}) = \sum_{k,l} \bm{\Delta h}_{k}^\intercal \;\; \bm{U}_l \;\; \bm{\Delta h}_{k-l}
\label{eq:distTRBMmatrix}
}
with
\eq{
\bm{U}_l = \sum_{d,d'=0}^{D-1} \bm{W}_d \bm{C}_{d-d'-l} \bm{W}_{d'}^\intercal
}
where $\bm{C}_d$ is the cross-covariance between neural responses with delay $d$.

In the special case of a TRBM with no interaction between different time bins ($D=1$), TRBM and RBM metrics are equivalent. 

\subsubsection{Continuous TRBM metric}
In order to define a metric based on the TRBM that does not require to binarize responses, we introduce a continuous time approximation of the semantic TRBM metric. 

$\bm{\Delta h}$ is the difference between $\<\bm{h}|\bm{\s}^{(1)}\>$ and  $\<\bm{h}|\bm{\s}^{(2)}\>$, and eq.~\eqref{eq:distTRBMmatrix} measures a norm of this difference. In order to express the TRBM-based metric in a form that is convenient for expression in continuous time, we approximate this difference using a linear expansion of the sigmoid function in eq.~\eqref{eq:hprobaTRBM}: 
\eq{
\Delta \bm{h}_k \approx &= \dfrac{1}{4} \sum_{d=0}^{D-1} \bm{W}_{d} \; \bm{\Delta \s}_{k-d}
}
where $\bm{\Delta \s} = \bm{\s}^{(1)} - \bm{\s}^{(2)}$. The semantic TRBM metric becomes:
\eq{
d (\bm{\s}^{(1)},\bm{\s}^{(2)}) = \sum_{k,l} \bm{\Delta \s}_k^\intercal \;\; \bm{V}_{l} \;\; \bm{\Delta \s}_{k-l}
}
where
\eq{
\bm{V}_l = \sum_{d,d'} \bm{Z}_d^\intercal  \;\; \bm{C}_{d'-d-l}  \;\; \bm{Z}_{d'}
}
and
\eq{
\bm{Z}_d = \sum_{d'} \bm{W}_{d'}^\intercal \; \bm{W}_{d+d'}
}

We dropped the 1/4 factor in $\bm{\Delta h}$, as multiplying a metric by a constant has no effect on its discriminating properties. 
This can be approximated in continuous time by the Euclidean metric corresponding to the following scalar product \citep{Paiva2010, Naud2011}:
\eq{
&\< R^{(1)}, R^{(2)} \> = \sum_{i,i',n,n'} \tilde{V}_{i,i'}(t^{(1)}_{in} - t^{(2)}_{i'n'})
\label{eq:contTRBMprod}
}
where $\bm{\tilde{V}}$ is a continuous time approximation for $\bm{V}$: $\bm{\tilde{V}}(l\Delta_t) = \bm{V}_{l}$ for any integer $l$. We used a piecewise linear interpolation for remaining times. The continuous TRBM metric is then:
\eq{
\begin{split}
& d_\text{cTRBM}(R^{(1)}, R^{(2)})^2 = \\
&\qquad \< R^{(1)}\!, R^{(1)} \>\! +\! \< R^{(2)}\!, R^{(2)} \> \!-\! 2 \< R^{(1)}\!, R^{(2)} \>.\\
\end{split}
\label{eq:contTRBM}
}

\subsection{Linear discriminability}
The linear discriminability is a measure that is independent of any metric, obtained by projecting responses onto a single direction. 
We measured binned responses $\bm{\s}_\text{ref}$ to multiple repetitions of the reference stimulus, and responses $\bm{\s}_{S\text{max}}$ to multiple repetitions of the largest amplitude of the same perturbation shape (typically 110 \textmu m). 
We computed the mean response to the reference, $\< \bm{\s}_\text{ref}\>$, and to the largest amplitude perturbation, $\<\bm{\s}_{S\text{max}}\>$, and projected all responses onto their difference:
we call $x_\text{ref} = (\<\bm{\s}_{S\text{max}}\> - \<\bm{\s}_\text{ref}\>) \cdot \bm{\s}_\text{ref}$ the projection of a response to the reference,
 and $x_S = (\<\bm{\s}_{S\text{max}}\> - \<\bm{\s}_\text{ref}\>) \cdot \bm{\s}_S$ the projection of a response to $S$ (when doing the projection, we recalculated the mean response by excluding the response that was being projected, to avoid over-fitting). The linear discriminability of $\bm{\s}_S$ is defined as the probability that $x_\text{ref} < x_S$. 

Note that although this definition of discriminability is convenient because it doesn't make any assumption about a metric. It is supervised as it requires us to know the mean response to a perturbation. Conversely, discriminability based on metrics can be computed for a single response to a perturbation.

During the experiment, to identify the range of perturbations that were neither too easy nor too hard to discriminate, we adapted perturbation amplitudes online using the Accelerated Stochastic Approximation algorithm \citep{Kesten58}  so that the linear discriminability converged to target value 85\%.

In order to compare metrics, we formed 3 groups of responses based on their linear discriminability: low (lower than 0.95), medium (higher of equal to 0.95 and lower than 1) and high (1).\\

\subsection{Code availability}

{The code for learning the models, computing their statistics and implementing the RBM and TRBM metrics is freely available at \url{https://github.com/ChrisGll/RBM_TRBM}.}

{\bf Ackownledgements.} 
We thank David Schwab for insightful discussions on RBMs. This work was supported by ANR TRAJECTORY, ANR OPTIMA, ANR IRREVERSIBLE (ANR-17-ERC2-0025-01), the French State program Investissements d'Avenir managed by the Agence Nationale de la Recherche [LIFESENSES: ANR- 10-LABX-65], European Commission grant from the Human Brain Project n. FP7-604102, and National Institutes of Health grant n. U01NS090501.

\bibliographystyle{pnas}

\end{document}